\documentclass{article}
\usepackage{frascatiphys}
\begin{document}
\title{ 
THE GZK FEATURE IN THE SPECTRUM\\
OF ULTRA HIGH ENERGY COSMIC RAYS
}
\author{
Daniel De Marco\\
{\em Universit\`a di Genova, Dipartimento di Fisica}\\
{\em \& INFN Sezione di Genova, Genova, ITALY}\\
}
\maketitle
\baselineskip=11.6pt
\begin{abstract}
The detection of the GZK feature in the cosmic ray spectrum, resulting
from the production of pions by ultra-high energy protons scattering off
the cosmic microwave background (CMB), can shed light on the mysterious
sources of these high energy particles.
Using numerical simulations we developed for the propagation of Ultra
High Energy Cosmic Rays (UHECR) in the CMB we determine the statistical
significance of the GZK feature in the spectrum of UHECR measured by
AGASA and HiRes and we show that, with the small sample of events with
energies $\sim 10^{20}$~eV detected thus far, an accurate and
statistically significant determination of the GZK feature is not
possible.
The data from these two experiments are best fit by two different
injection spectra in the region below $10^{20}$~eV, and a comparison of
the spectra suggests the presence of about a 30\% systematic errors in
the relative energy determination. Correcting for these systematics, the
two experiments are best fit by the same injection spectrum in the region
below $10^{20}$~eV, while above this threshold they maintain their
disagreement, but only at the 2$\sigma$ level. These results clearly show
the need for much larger experiments such as Auger, EUSO and OWL, that
can increase the number of detected events by one or two orders of
magnitude making the determination of the GZK feature feasible.
\end{abstract}
\baselineskip=14pt
\section{Introduction}
Astrophysical proton sources distributed homogeneously in the universe
produce a feature in the energy spectrum due to the production of pions
off the CMB. This feature, consisting of a rather sharp suppression of
the flux, occurs at energies above $E_{\rm GZK} \sim7 \times
10^{19}$~eV, and it is now known as the GZK cutoff. Almost forty years
after this prediction it is not yet clear if this effect is observed or
not due to the discrepancy between the results of the two largest
experiments measuring the spectrum of Ultra High Energy Cosmic Rays
(UHECRs). AGASA\cite{AGASA} reports a higher number of events above
$E_{\rm GZK}$ than expected while HiRes\cite{HiRes} reports a flux
consistent with the GZK feature.  Here we report on a detailed
investigation\cite{arti} of the statistical significance of this discrepancy as
well as the significance of the presence or absence of the GZK feature
in the data. We find that neither experiment has the necessary
statistics to establish if the spectrum of UHECRs has a GZK feature.  In
addition, a systematic error in the energy determination of the two
experiments seems to be required in order to make the two sets of
observations compatible in the low energy range, $10^{18.5}-10^{19.6}$
eV, where enough events have been detected to make the measurements
reliable. Taking into account the systematics, the two experiments
predict compatible fluxes at energies below $E_{\rm GZK}$ and at
energies above $E_{\rm GZK}$ the fluxes are within $\sim 2\sigma$ of
each other. 

The detection or non-detection of the GZK feature in the cosmic ray
spectrum remains open to investigation by future generation experiments,
such as the Pierre Auger project and the EUSO and OWL experiments.

The paper is organized as follows: in \S~2 we briefly describe
the simulations used to propagate cosmic rays, in \S~3 we illustrate the
present observational situation, limiting ourselves to AGASA and HiResI,
and compare the data to the predictions of our simulations. We conclude
in \S~4.

\section{Cosmic Rays Propagation}
We assume that UHECRs are protons injected in extragalactic sources with
a power-law spectrum with slope $\gamma$ and an exponential cutoff at
$E_{\rm max}=10^{21.5}\,{\rm eV}$, large enough not to affect the
statistics at much lower energies. Based on the results of
\cite{blanton} we assume a spatially uniform distribution of sources and
do not take into account luminosity evolution in order to avoid the
introduction of additional parameters.
We simulate the propagation of protons from source to observer by including the
photo-pion production, pair production, and adiabatic energy losses due to
the expansion of the universe.
In each step of the simulation, we calculate the pair production losses
using the continuous energy loss approximation given the small inelasticity in
pair production ($2 m_{\rm e}/m_{\rm p}\simeq10^{-3}$).  For the rate of
energy loss due to pair production at redshift $z=0$, $\beta_{\rm pp}(E,z=0)$,
we use the results from \cite{blumenthal,czs}. At a given redshift
$z>0$,
\begin{equation}
\beta_{\rm pp}(E,z)=(1+z)^3 \beta_{\rm pp}((1+z)E, z=0)\,.
\end{equation}
Similarly, the rate of adiabatic energy losses due to redshift is
calculated in each step  using
\begin{equation}
\beta_{\rm rsh}(E,z)=H_0 \left[\Omega_M (1+z)^3 +
\Omega_\Lambda\right]^{1/2}
\end{equation}
with $H_0=75 ~ {\rm km~ s^{-1} Mpc}^{-1}$.

The photo-pion production is simulated in a way similar to that described
in ref.~\cite{blanton}. In each step, we first calculate the
average number of photons able to interact via photo-pion production
through the expression:
\begin{equation}
\langle N_{\rm ph}(E,\Delta s) \rangle=\frac{\Delta s}{l(E, z)},
\end{equation}
where $l(E,z)$ is the interaction length for photo-pion production of a
proton with energy $E$ at redshift $z$ and $\Delta s$ is a step size,
chosen to be much smaller than the interaction length (typically we
choose $\Delta s=100 ~{\rm kpc}/(1+z)^3$).

We then sample a Poisson
distribution with mean $\langle N_{\rm ph}(E,\Delta s) \rangle$, to determine
the  actual number of photons encountered during the step $\Delta s$. When a
photo-pion interaction occurs, the energy $\epsilon$ of the photon is extracted
from the Planck distribution, $n_{ph}(\epsilon,T(z))$, with temperature
$T(z)=T_0 (1+z)$, where $T_0=2.728$ K is the temperature
of the cosmic microwave background at present. Since the microwave photons
are isotropically distributed, the interaction angle, $\theta$, between the
proton and the photon is sampled randomly from a distribution which is flat
in $\mu={\rm cos}\theta$. Clearly only the values of $\epsilon$ and
$\theta$ that generate a center of mass energy above the threshold for
pion production are considered. The energy of the proton in the final
state is calculated at each interaction from kinematics.
The simulation is carried out until the statistics of events
detected above some energy reproduces the experimental numbers.
By normalizing the simulated flux by the number of events above an energy
where experiments have high statistics, we can then ask what are the
fluctuations in numbers of events above a higher energy where
experimental results are sparse. The fits are therefore most sensitive
to the energy regions below $E_{GZK}$ and give a good estimate of the
uncertainties in the present experiments for energies above  $E_{GZK}$.
In this way we have a direct handle on the fluctuations that can be
expected in the observed flux due to the stochastic nature of photo-pion
production and to cosmic variance.

The simulation proceeds in the following way: a source distance is
generated at random from a uniform distribution in a universe with
$\Omega_\Lambda=0.7$ and $\Omega_m=0.3$, a particle energy is assigned
from a distribution that reflects the injection spectrum then this
particle is propagated to the observer and its energy recorded. This
procedure is repeated until the number of events above a threshold
energy, $E_{th}$ is reproduced. With this procedure we can assess the
significance of results from present experiments with limited statistics
of events. There is an additional complication in that the aperture of
the experiment usually depends on energy. This is taken into account by
allowing the event to be detected or not depending upon the function
$H(E)$ that describes the energy dependence of the aperture.

We only study the spectrum above $10^{18.5}$ eV where the flux is
supposed to be dominated by  extragalactic sources. For this energy
range, we focus on the experiments that have the best statistics: AGASA
and HiResI.  For AGASA data, the simulation is stopped when the number
of events above $E_{th} = 10^{19}$ eV equals 866. For HiRes this number
is 300. The statistical error in the energy determination is accounted
for in our simulation by generating a {\it detection} energy chosen at
random from a Gaussian distribution centered at the arrival energy $E$
and with width $\Delta E/E=30\%$ for both experiments.

Our simulations reproduce well the predictions of analytical calculations,
in particular at the energies where energy losses may be approximated as
continuous (see \cite{arti} for a detailed comparison).
\section{AGASA versus HiResI}
\begin{figure}[t]
\vspace{4.5cm}
\includegraphics{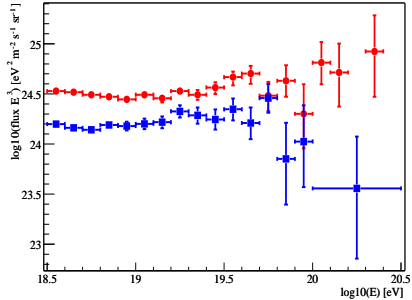}
\caption{\it Circles show the AGASA spectrum while squares show the
HiResI spectrum. \label{fig:agasahires} }
\end{figure}

The two largest experiments that measured the flux of UHECRs report
apparently conflicting results. The data of AGASA and HiResI on the
flux of UHECRs multiplied as usual by the third power of the energy
are plotted in Fig. \ref{fig:agasahires} (the squares are the HiResI
data while the circles are the AGASA data).
The figure shows that HiResI data are systematically below AGASA data
and that HiResI sees a suppression at $\sim 10^{20}$ eV that resembles
the GZK feature while AGASA does not.

We apply our simulations here to the statistics of events of both AGASA
and HiResI in order to understand whether the discrepancy is
statistically significant and whether the GZK feature has indeed been
detected in the cosmic ray spectrum.
In order to understand the difference, if any, between AGASA and HiRes
data we first determine the injection spectrum required to best fit the
observations.
The injection spectrum is taken to be a power law with index $\gamma$
between 2.3 and 2.9 with steps of 0.1. For each injection spectrum we
calculated the $\chi^2$ indicator (averaged over 400 realizations for
each injection spectrum). The errors used for the evaluation of the
$\chi^2$ are due to the square roots of the number of observed events.
It is important to note that the fits are dominated by the low energy
data rather than by the poorer statistics at the higher energies.

If the data at energies above $10^{18.5}$ eV are taken into account
for both experiments, the best fit spectra are $E^{-2.8}$ for AGASA and
$E^{-2.6}$ for HiRes. If the data at energies above $10^{19}$ eV are
used for the fit, the best fit injection spectrum is $E^{-2.6}$ for
AGASA and between $E^{-2.7}$ and $E^{-2.8}$ for HiRes. If the fit is
carried out on the highest energy data ($E>10^{19.6}$ eV), AGASA
prefers an injection spectrum between $E^{-2.5}$ and $E^{-2.6}$, while
$E^{-2.8}$ or $E^{-2.9}$ fit better the HiRes data in the same energy
region. Note that the two sets of data uncorrected for any possible
systematic errors require different injection spectra that change with
$E_{th}$.

In order to quantify the significance of the detection or lack of the
GZK flux suppression, we calculate the mean number of events above two
energy thresholds ($10^{19.6}$~eV and $10^{20}$~eV), 
$\langle N (E > E_{th}) \rangle$, for different injection spectra and
compare them to the experimental results. We find that while HiResI is
consistent with the existence of the GZK feature in the spectrum of
UHECRs, AGASA detects an increase in flux, but only at the $\sim
2.1\sigma_{\rm tot}$ level. Here $\sigma_{tot}=(\sigma_{sim}^2+
\sigma_{obs}^2)^{1/2}$ is the combined uncertainty from simulations and
observed data.

\begin{figure}[t]
\vspace{4.5cm}
\includegraphics{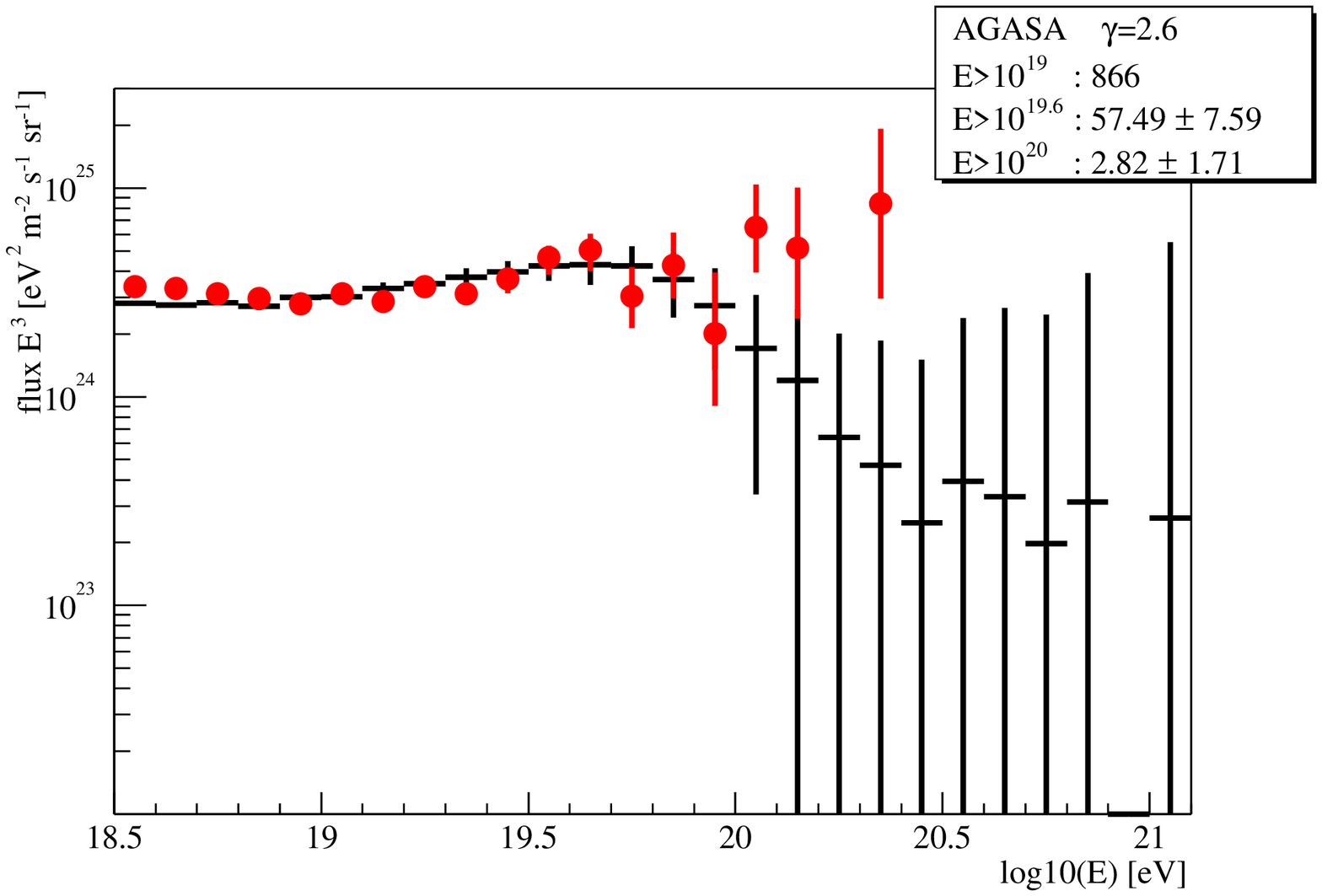}
\includegraphics{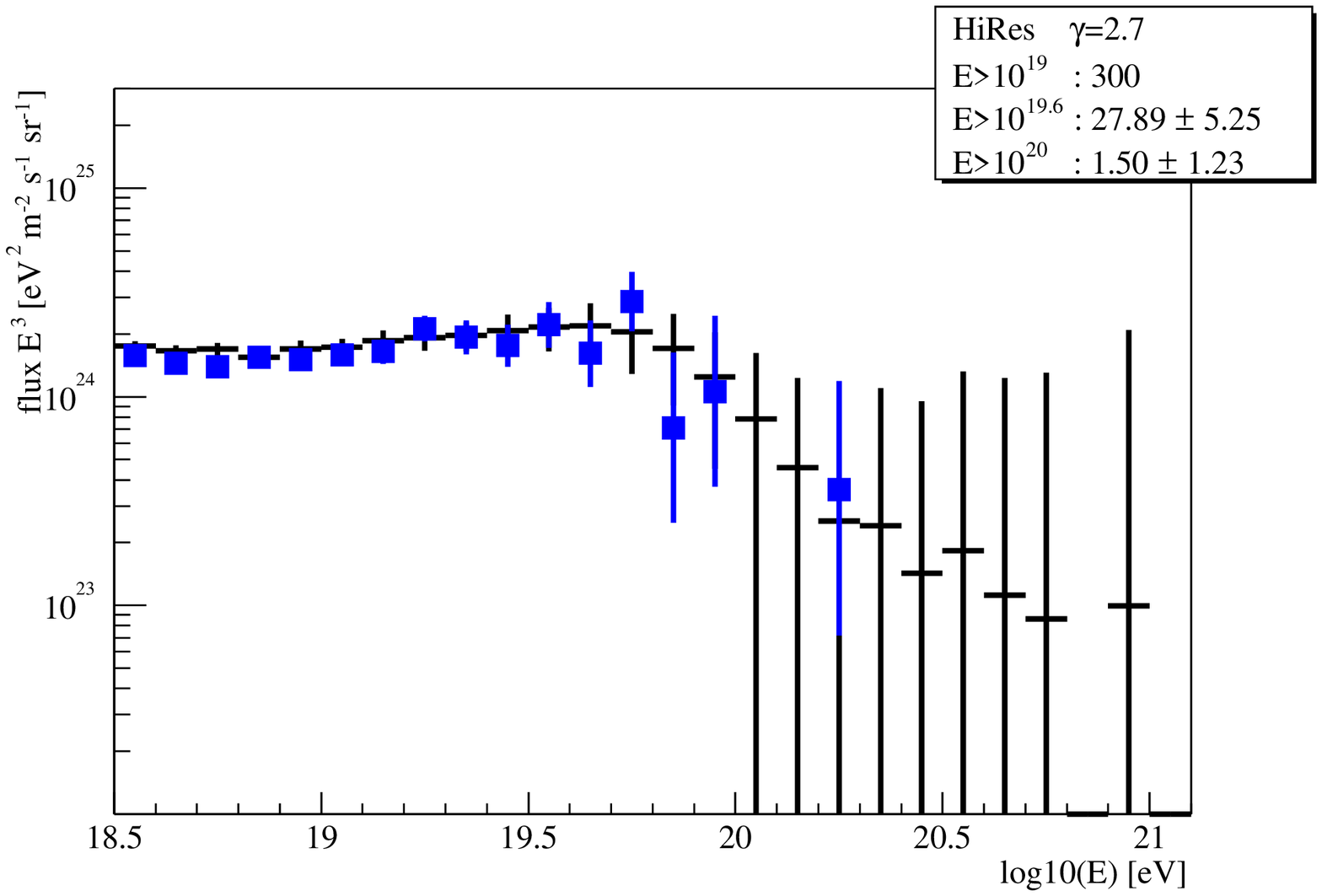}
\caption{\it Simulations of AGASA and HiRes statistics. crosses with
error bars: simulations, grey points: experimental data.
\label{uncorrected}}
\end{figure}

A more graphical representation of the uncertainties involved are
displayed in Fig. \ref{uncorrected}. These plots show clearly the low
level of significance that the detections above $E_{GZK}$ have with low
statistics.  The large error bars that are generated by our simulations
at the high energy end of the spectrum are mainly due to the stochastic
nature of the process of photo-pion production: in some realizations
some energy bins are populated by a few events, while in others the few
particles in the same energy bin do not produce a pion and get to the
observer unaffected. The large fluctuations are unavoidable with the
extremely small statistics available with present experiments. On the
other hand, the error bars at lower energies are minuscule, so that the
two data sets (AGASA and HiResI) cannot be considered to be two
different realizations of the same phenomenon. Instead, systematic
errors in at least one if not both experiments are needed to explain the
discrepancies at lower energies.

As seen in Fig. \ref{fig:agasahires}, the difference between the AGASA
and HiResI spectra is not only in the presence or absence of the GZK
feature: the two spectra, when multiplied by $E^3$, are systematically
shifted by about a factor of two. This shift suggests that there may be
a systematic error either in the energy or the flux determination of at
least one of the two experiments. Possible systematic effects have been
discussed in \cite{AGASA2} for the AGASA collaboration and in
\cite{HiRes2} for HiResI. A systematic error of $\sim 15\%$ in the
energy determination is well within the limits that are allowed by the
analysis of systematic errors carried out by both collaborations.

In order to illustrate the difficulty in determining the existence of
the GZK feature in the observed data in the presence of systematic
errors, we split the energy gap by assuming that the energies as
determined by the AGASA collaboration are overestimated by $15\%$, while
the HiRes energies are underestimated by the same factor. In this case
the total number of events above $10^{19}$ eV is reduced for AGASA from
866 to 651, while for HiResI it is enhanced from 300 to 367. We ran our
simulations with these new numbers of events and repeat the statistical
analysis described above.  

For AGASA, the best fit injection spectrum is now between $E^{-2.5}$ and
$E^{-2.6}$ above $10^{19}$ eV and above $10^{19.6}$ eV. For the HiRes
data, the best fit injection spectrum is $E^{-2.6}$ for the whole set of
data, independent of the threshold. It is interesting to note that the
best fit injection spectrum appears much more stable after the
correction of the $15\%$ systematics has been carried out. Moreover, the
best fit injection spectra as derived for each experiment independently
coincides for the corrected data unlike the uncorrected case. This
suggests that combined systematic errors in the energy determination at
the $\sim$ 30\% level may in fact be present.

As above in order to quantify the significance of the detection or lack
of the GZK flux suppression, we calculate the mean number of events
above two energy thresholds ($10^{19.6}$~eV and $10^{20}$~eV), $\langle N
(E > E_{th}) \rangle$, for different injection spectra and compare them
to the experimental results. We find that HiResI continues to be
consistent with the existence of the GZK feature, while the discrepancy
of the AGASA data is reduced at the level of $1.5\sigma_{\rm tot}$.

\begin{figure}[t]
\vspace{4.5cm}
\includegraphics{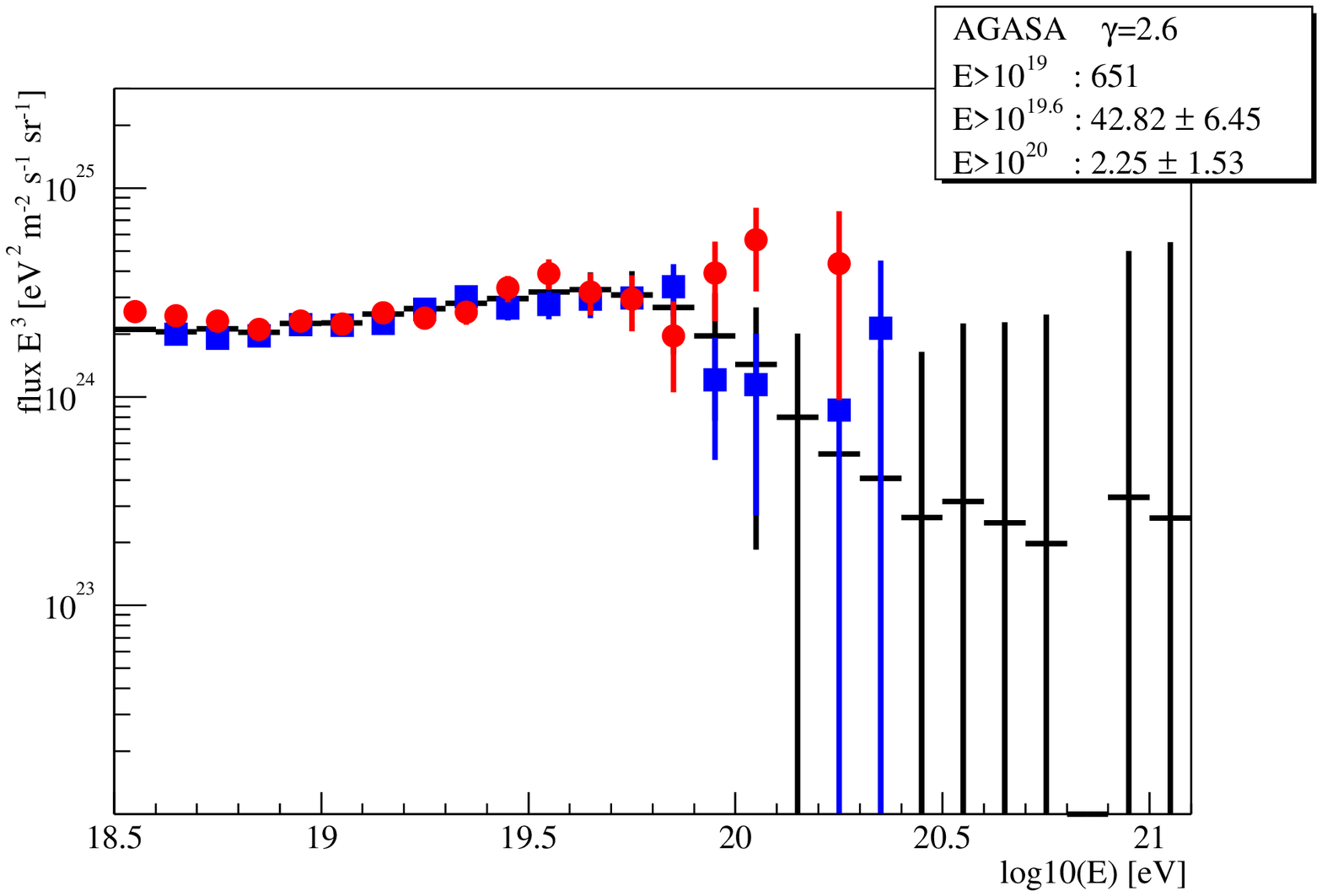}
\includegraphics{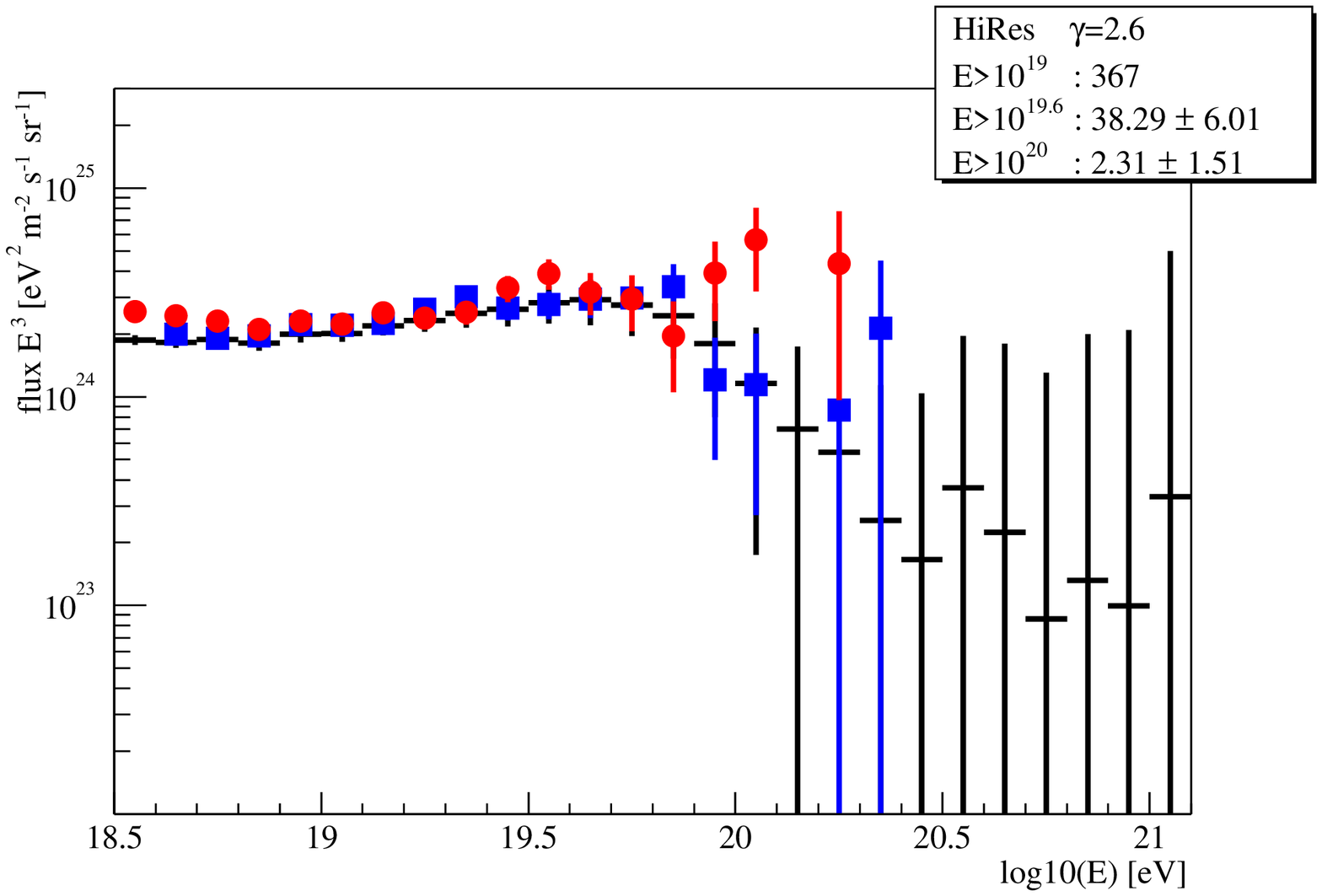}
\caption{\it Simulated spectra for the best fit injection spectrum with
$\gamma=2.6$. Left panel shows simulations for the AGASA event
statistics after correcting for energy by an overall shift of -15\%. The
right panel shows  the fluctuations expected for the event statistics of
HiResI after shifting HiResI energies by +15\%. The shifted data for
AGASA (grey circles) and  HiResI (dark squares) are shown in both
panels. \label{corrected}}
\end{figure}

In Fig. \ref{corrected}, we plot the simulated spectra for injection
spectrum $E^{-2.6}$ and compare them to the corrected data of AGASA (left
plot) and HiResI (right plot).

\section{Conclusions}
We considered the statistical significance of the UHECR spectra measured
by the two largest experiments in operation, AGASA and HiRes. The
spectrum released by the HiResI collaboration  seems to suggest the
presence of a GZK feature. This has generated claims that the GZK cutoff
has been detected, reinforced by data from older experiments
\cite{waxmanb}. However, no evidence for such a feature has been  found
in the AGASA experiment.  We compared the data with theoretical
predictions for the propagation of UHECRs on cosmological distances with
the  help of numerical simulations. We find that the very low statistics
of the presently available data hinders any statistically significant
claim for either detection or the lack of the GZK feature.

A comparison of the spectra obtained from AGASA and
HiResI shows a systematic shift of the two data sets, which may be
interpreted as a systematic error in the relative energy determination
of about $30\%$. If no correction for this systematic shift is carried
out, the AGASA and HiResI data sets are best fit by two different
injection spectra that depend on the threshold used for the fit.
With the best fits to the injection spectrum the
AGASA data depart from the prediction of a GZK feature by $2.3\sigma$
for $\gamma=2.6$. The HiRes data
are fully compatible with the prediction of a GZK feature in the
cosmic ray spectrum. 
It is clear that, if confirmed by future experiments with
much larger statistics, the increase in flux  relative to the GZK
prediction hinted by AGASA would be of great interest. This may signal the
presence of a new component at the highest energies that compensates for the
expected suppression due to photo-pion production, or the effect of new
physics in particle interactions (for instance the violation of Lorentz
invariance or new neutrino interactions).

Identifying the cause of the systematic energy and/or flux shift between
the AGASA and the HiRes spectra is crucial for understanding the nature
of UHECRs. This discrepancy has stimulated a number of efforts  to search
for the source of these systematic errors including the construction of
hybrid detectors, such as Auger, that utilize both ground arrays and
fluorescence detectors. A possible overestimate of the AGASA energies by
$15\%$ and a corresponding underestimate of the HiRes energies by the same
amount would in fact bring the two data sets in agreement in the
region of energies below $10^{20}$~eV. In this case both experiments are
consistent with a GZK feature with large error bars. The AGASA excess
is at the level of  $1.5\sigma$. Interestingly
enough, the correction by $15\%$ in the error determination implies
that the best fit injection spectrum becomes basically the same
for both experiments ($E^{-2.6}$).

\begin{figure}[t]
\vspace{4.5cm}
\includegraphics{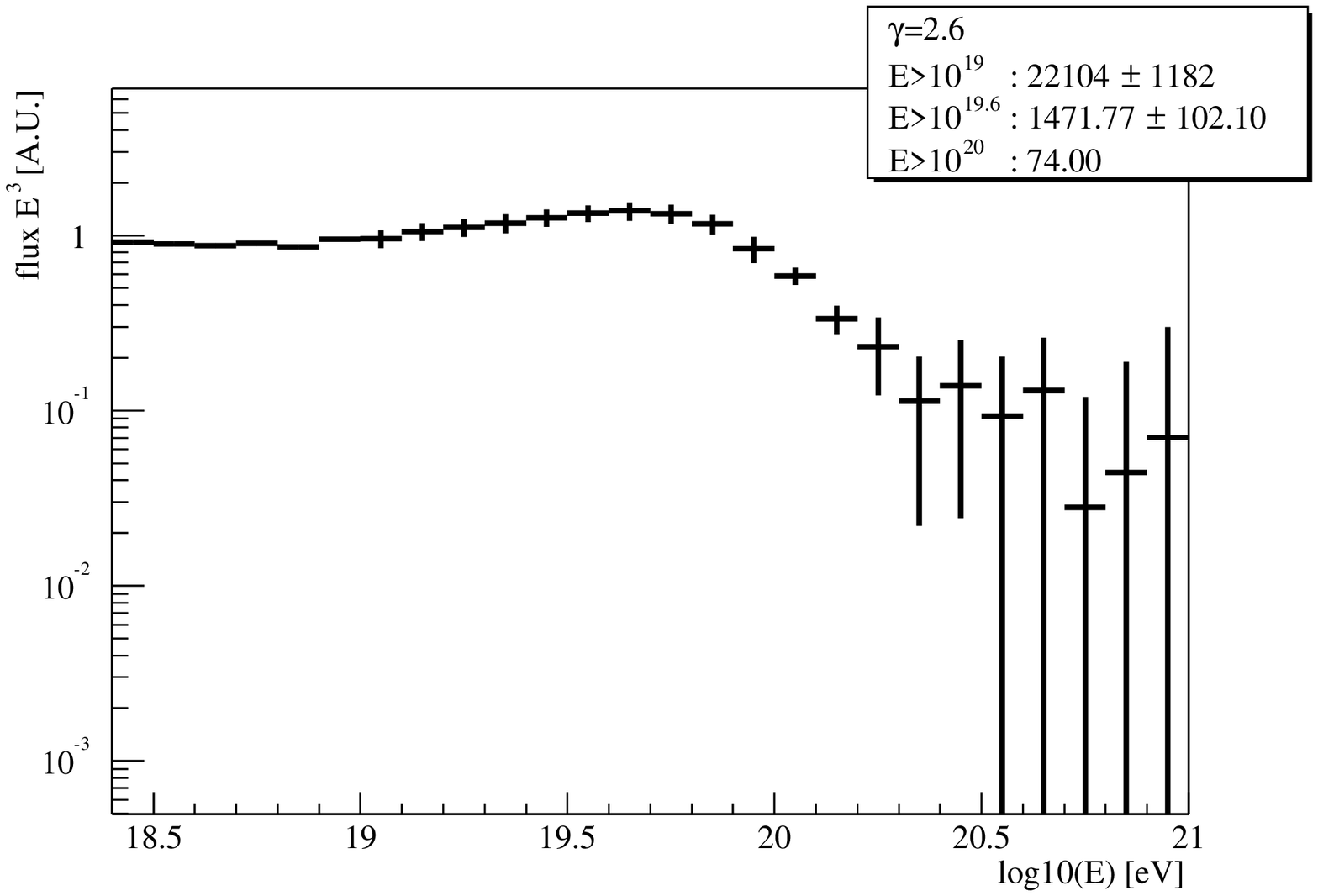}
\includegraphics{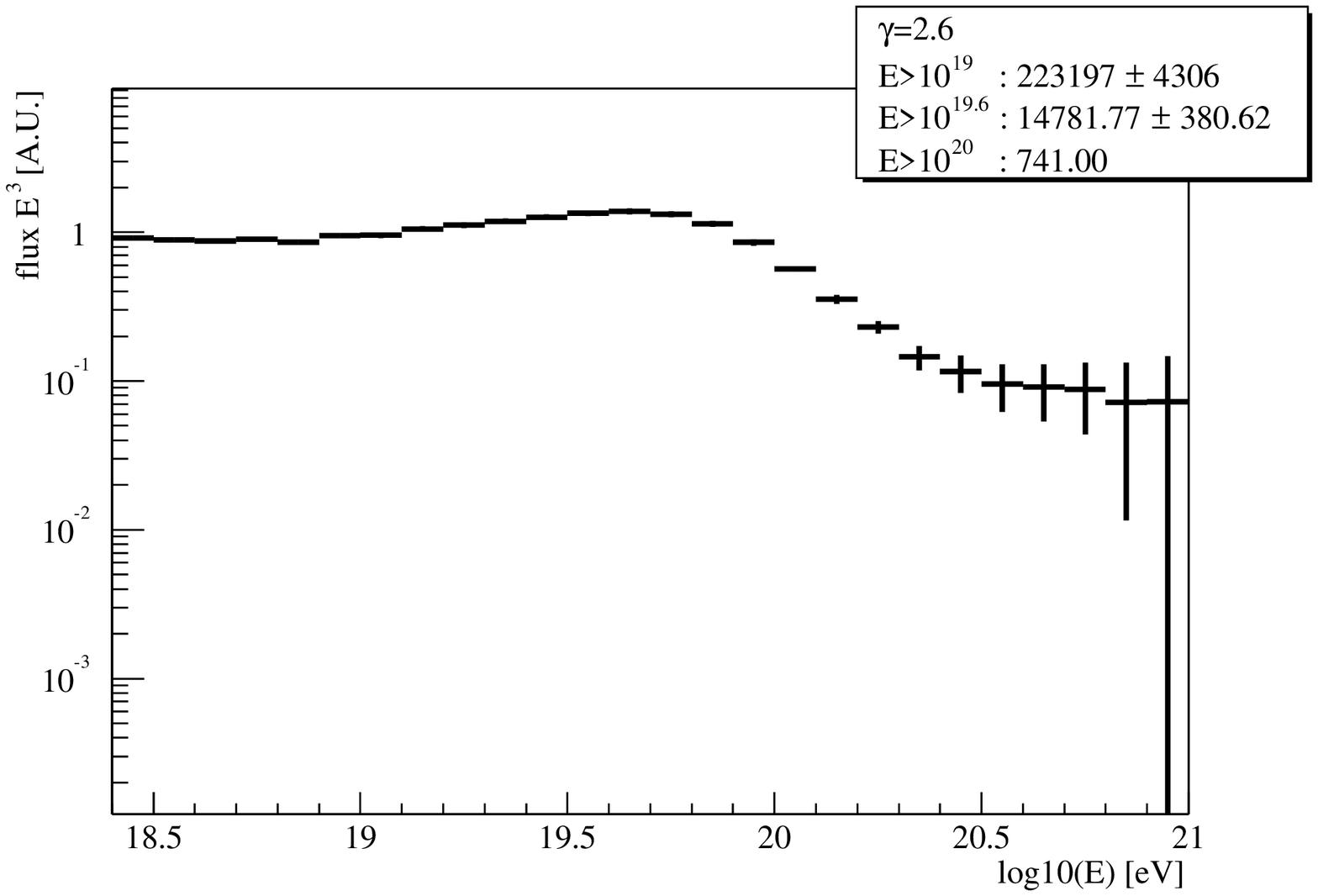}
\caption{\it Predicted spectra and error bars for 3 years of operation
of Auger (left plot) and EUSO (right plot).
\label{predictions}}
\end{figure}

With the low statistical significance of either the excess flux seen by
AGASA or the discrepancies between AGASA and HiResI, it is inaccurate to
claim either the detection of the GZK feature or the extension of the
UHECR spectrum beyond $E_{GZK}$ at this point in time. A new generation
of experiments is needed to finally give a clear answer to
this question. In Fig. \ref{predictions} we report the simulated spectra
that should be achieved in 3 years of operation of Auger (left panel)
and EUSO (right panel). The error bars reflect the fluctuations expected
in these high statistics experiments for the case of injection spectrum
$E^{-2.6}$. (Note that the energy threshold for detection by EUSO is not
yet clear). It is clear that the energy region where statistical
fluctuations dominate the spectrum is moved to $\sim 10^{20.6}$ eV for
Auger, allowing a clear identification of the GZK feature. The {\it
fluctuations} dominated region stands beyond $10^{21}$ eV for EUSO.

\end{document}